\documentclass[preprint,preprintnumbers, prd, floatfix, superscriptaddress,nofootinbib] {revtex4-1}
\usepackage{epsfig}
\usepackage{subfigure}
\usepackage{dcolumn}
\usepackage{bm}
\usepackage[usenames ,dvipsnames]{xcolor}
\usepackage{slashed}
\usepackage{graphicx,color}
\newcommand{\strich}[1]{#1  \! \! \slash}

\begin{document}
\title{Fragmentation fraction $f_{\Omega_b}$ and the $\Omega_b\to \Omega J/\psi$ decay\\
in the light-front formalism}

\author{Yu-Kuo Hsiao}
\email{yukuohsiao@gmail.com}
\affiliation{School of Physics and Information Engineering, Shanxi Normal University, Taiyuan, 030031}

\author{Chong-Chung Lih}
\email{cclih@phys.nthu.edu.tw}
\affiliation{Department of Optometry, Central Taiwan University of Science and Technology, Taichung 40601}

\date{\today}

\begin{abstract}
One has measured 
$f_{\Omega_b}{\cal B}(\Omega_b^-\to \Omega^- J/\Psi)$ at the level of $10^{-6}$,
where the fragmentation faction $f_{\Omega_b}$ 
is to evaluate the $b$-quark to $\Omega_b^-$ production rate.
Using the $\Omega_b\to \Omega$ transition form factors 
calculated in the light-front quark model,
we predict ${\cal B}(\Omega_b^-\to \Omega^- J/\Psi)
=(5.3^{+3.3+3.8}_{-2.1-2.7})\times 10^{-4}$. In particular, 
we extract $f_{\Omega_b}=(0.54^{+0.34+0.39+0.21}_{-0.22-0.28-0.15})\times 10^{-2}$,
demonstrating that the $b$ to $\Omega_b$ productions
are much more difficult than the $b$ to $\Lambda_b(\Xi_b)$ ones.
Since $f_{\Omega_b}$ has not been determined experimentally,  
$f_{\Omega_b}$ added to theoretical branching fractions 
can be compared to future measurements of the $\Omega_b$ decays.
\end{abstract}

\maketitle
\section{introduction}
The anti-triplet $b$-baryons $(\Lambda_b,\Xi_b^0,\Xi_b^-)$ 
and $\Omega_b^-$ all decay weakly~\cite{pdg}, 
where $\Omega_b$ belongs to the sextet $b$-baryon states. 
Interestingly, only $\Omega_b$ is allowed 
to have a direct transition to ${\bf B}^*$ in the weak interaction,
where ${\bf B}^*$ stands for a spin-3/2 decuplet baryon.
This is due to the fact that $\Omega_b$ and ${\bf B}^*$
both have totally symmetric quark orderings. 
By contrast, the anti-triplet baryon ${\bf B}_b$ consisting of $(q_1q_2-q_2q_1)b$
mismatches ${\bf B}^*$ with $(q_1 q_2+q_2q_1)q_3$ in the ${\bf B}_b$ to ${\bf B}^*$ transition. 
Clearly, the $\Omega_b$ decay into ${\bf B}^*$ worths an investigation.

One has barely measured the $\Omega_b$ decays.
Moreover, the fragmentation fraction $f_{{\bf B}_b(\Omega_b)}$
that evaluates the $b$-quark to ${\bf B}_b(\Omega_b)$ production rate
has not been determined yet. Consequently,
the charmful $\Omega_b$ decay channel
$\Omega_b^-\to \Omega^- J/\Psi$ can only be partially measured.
In addition to $\Lambda_b\to \Lambda J/\psi$ and $\Xi_b^-\to \Xi^- J/\psi$,
the partial branching fractions are given by~\cite{pdg}
\begin{eqnarray}\label{data1}
f_{\Omega_b}{\cal B}(\Omega_b^-\to \Omega^- J/\Psi)
&=&(2.9^{+1.1}_{-0.8})\times 10^{-6}\,,\nonumber\\
f_{\Lambda_b}{\cal B}(\Lambda_b\to \Lambda J/\psi )
&=&(5.8\pm 0.8)\times 10^{-5}\,,\nonumber\\
f_{\Xi_b}{\cal B}(\Xi_b^-\to \Xi^- J/\psi)
&=&(1.02^{+0.26}_{-0.21})\times 10^{-5}\,,
\end{eqnarray} 
where $f_{\Xi_b}=f_{\Xi_b^{-(0)}}$.
Some theoretical attempts have been given to extract 
$f_{{\bf B}_b(\Omega_b)}$~\cite{Hsiao:2015cda,Hsiao:2015txa,Jiang:2018iqa}.
Using the calculations of ${\cal B}(\Lambda_b\to \Lambda J/\psi )$ and 
${\cal B}(\Xi_b^-\to \Xi^- J/\psi)$~\cite{Hsiao:2015cda,Hsiao:2015txa},
one extracts $f_{\Lambda_b}$ and $f_{\Xi_b}$ as some certain numbers.
Without a careful study of $\Omega_b^-\to \Omega^- J/\Psi$~\cite{Hsiao:2015cda,Hsiao:2015txa},
it is roughly estimated that $f_{\Omega_b}<0.108$. Therefore,
it can be an important task to explore the charmful $\Omega_b^-\to \Omega^- J/\Psi$ decay.

See Fig.~\ref{fig1}, $\Omega_b^-\to\Omega^- J/\Psi$ is depicted  
to proceed through the $\Omega_b^-\to\Omega^-$ transition, while
$J/\Psi$ is produced from the internal $W$-boson emission. 
To calculate the branching fraction, the information of 
the $\Omega_b\to\Omega$ transition is required.
On the other hand, the light-front quark model has provided
its calculation on the $\Omega_c\to\Omega$ transition form factors,
such that one interprets the relative branching fractions of 
$\Omega_c^0\to \Omega^-\rho^+$ and $\Omega_c^0\to \Omega^- \ell^+\bar \nu_\ell$
to that of $\Omega^-\pi^+$~\cite{Hsiao:2020gtc}.
Therefore, we propose to calculate the $\Omega_b^-\to \Omega^-$ transition form factors
in the light-front formalism, as applied to the $\Omega_c$ decays
as well as the other heavy hadron decays~\cite{Zhao:2018zcb,
Bakker:2003up,Ji:2000rd,Bakker:2002aw,Choi:2013ira,
Cheng:2003sm,Schlumpf:1992vq,Hsiao:2019wyd,Jaus:1991cy,
Melosh:1974cu,Dosch:1988hu,Zhao:2018mrg,Geng:2013yfa,Geng:2000if,
Ke:2012wa,Ke:2017eqo,Ke:2019smy,Hu:2020mxk,Chung:1988mu}.
We will be able to predict ${\cal B}(\Omega_b^-\to\Omega^- J/\Psi)$, and
extract $f_{\Omega_b}$. Besides, we will compare
the branching fractions of $\Omega_b^-\to \Omega^- J/\Psi$,
$\Lambda_b\to \Lambda J/\psi$ and $\Xi_b^-\to \Xi^- J/\psi$,
and their fragmentation fractions.

\newpage
\section{Formalism}
%
\begin{figure}[t]
\centering
\includegraphics[width=3.4 in]{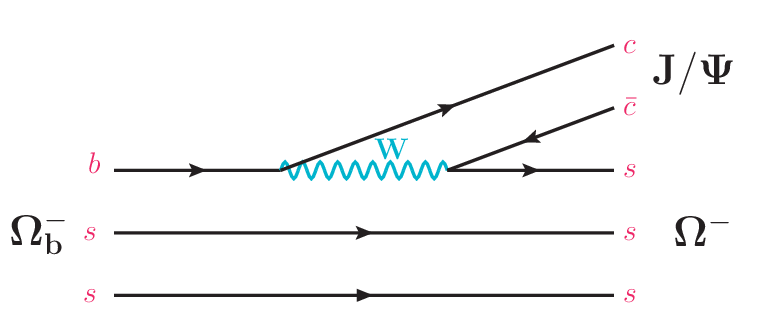}
\caption{Feynman diagram for $\Omega_b^-\to\Omega^- J/\Psi$.}\label{fig1}
\end{figure}
%
According to Fig.~\ref{fig1}, the amplitude of $\Omega_b^-\to\Omega^- J/\Psi$  
combines the matrix elements of 
the $\Omega_b^-\to\Omega^-$ transition and $J/\Psi$ production,
written as~\cite{Hsiao:2015cda,Hsiao:2015txa}
\begin{eqnarray}\label{amp1}
{\cal M}(\Omega_b^-\to\Omega^- J/\Psi)&=&
\frac{G_F}{\sqrt 2}V_{cb}V_{cs}^*a_2\,
\langle J/\psi|\bar c\gamma^\mu(1- \gamma_5) c|0\rangle
\langle\Omega^-|\bar s\gamma_\mu(1-\gamma_5) b|\Omega_b^-\rangle\,, 
\end{eqnarray}
where $G_F$ is the Fermi constant, and $V_{cb(s)}^{(*)}$ 
the Cabibbo-Kobayashi-Maskawa (CKM) matrix element.
The factorization derives that $a_2=c_2^{eff}+c_1^{eff}/N_c$, 
where $c_{1,2}^{eff}$ are the effective Wilson coefficients, and 
$N_c$ the color number~\cite{ali,Hsiao:2014mua}. For the $J/\Psi$ production,
the matrix elements read~\cite{Becirevic:2013bsa} 
\begin{eqnarray}\label{Jdecay}
\langle J/\psi|\bar c\gamma^\mu(1- \gamma_5) c|0\rangle
=m_{J/\psi} f_{J/\psi}\varepsilon_\mu^*\,,
\end{eqnarray}
where $m_{J/\psi}$, $f_{J/\psi}$ and $\varepsilon_\mu^*$ are 
the mass,  decay constant and  polarization four-vector, respectively.
The matrix elements of the $\Omega_b^-(bss)\to\Omega^-(sss)$ transition
are parameterized as~\cite{Zhao:2018mrg,Gutsche:2018utw}
\begin{eqnarray}\label{transitionVA1}
&&
\langle T^\mu\rangle\equiv 
\langle\Omega(sss)|\bar s\gamma^\mu(1-\gamma_5) b|\Omega_b(bss)\rangle\nonumber\\
&& =  \bar{u}_{\alpha}
\left[\frac{P^{\alpha}}{M}\left(\gamma^{\mu}F^V_{1}
+\frac{P^{\mu}}{M} F^V_{2}
+\frac{P^{\,\prime\mu}}{M^{\prime}}F^V_{3}\right)+g^{\alpha\mu}F^V_{4}\right]
\gamma_{5}u\nonumber\\
&&\quad-\bar{u}_{\alpha}
\left[\frac{P^{\alpha}}{M}\left(\gamma^{\mu}F^A_{1}
+\frac{P^{\mu} }{M}F^A_{2}
+\frac{P^{\,\prime\mu}}{M^{\prime}}F^A_{3}\right)
+g^{\alpha\mu}F^A_{4}\right]u\,,
\end{eqnarray}
where $M^{(\prime)}$ and $P^{(\prime)}$ represent the mass and momentum of
$\Omega_b(\Omega)$, respectively, and 
$F^{V,A}_i$ ($i=1,2, ..,4$) are the form factors.
By substituting the matrix elements of 
Eqs.~(\ref{Jdecay}, \ref{transitionVA1}) for those of Eq.~(\ref{amp1}), 
we derive the amplitude in the helicity basis~\cite{Gutsche:2018utw},
\begin{eqnarray}
{\cal M}&=&c_W \sum_{\lambda_\Omega,\lambda_J} 
(H^V_{\lambda_\Omega \lambda_J}-H^A_{\lambda_\Omega \lambda_J})\,,
\end{eqnarray}
where $c_W\equiv (G_F/\sqrt 2) V_{cb}V_{cs}^*\,a_2 m_{J/\psi} f_{J/\psi}$, and
$\lambda_\Omega=(\pm 3/2,\pm 1/2)$ and $\lambda_J=(0,\pm 1)$
denote the helicity states of $\Omega$ and $J/\Psi$, respectively.
Due to the helicity conservation, 
$\lambda_{\Omega_b}=\lambda_\Omega-\lambda_J$ should be respected,
where $\lambda_{\Omega_b}=\pm 1/2$. Subsequently, 
we obtain~\cite{Gutsche:2018utw}
\begin{eqnarray}\label{HVA}
&&
H_{\frac120}^{V(A)}= \sqrt{\frac{2}{3}\frac{Q^2_\mp}{q^2}}
\left[ F_1^{V(A)} \left(\frac{Q^2_\pm M_\mp}{2MM'}\right)
\mp\left(F_2^{V(A)}+F_3^{V(A)}\frac{M}{M'}\right) \left(\frac{|\vec{P}'|^2}{M'}\right)
\mp F_4^{V(A)}\bar M'_- \right]\,,\nonumber\\
&&
H_{\frac121}^{V(A)}=-\sqrt{\frac{Q^2_\mp}{3}}
\left[F_1^{V(A)} \left(\frac{Q^2_\pm}{M M'}\right) -F_4^{V(A)}\right]\,,\nonumber\\
&&
H_{\frac321}^{V(A)} = \mp \sqrt{Q^2_\mp} \, F_4^{V(A)}\,,
\end{eqnarray}
and $H^{V(A)}_{-\lambda_\Omega -\lambda_J} = \mp H^{V(A)}_{\lambda_\Omega\lambda_f}$,
with $M_\pm = M\pm M'$, $Q^2_\pm = M_\pm^2 - q^2$, 
$\bar M_{\pm}^{(\prime)}=(M_+M_-\pm q^2)/(2M^{(\prime)})$ and
$|\vec{P}'|=\sqrt{Q^2_+ Q^2_-}/(2M)$.

In the light-front quark model, we can calculate the form factors.
To start with, we consider the baryon as a bound state 
that consists of three quarks $q_1$, $q_2$ and $q_3$, 
where $q_{2,3}$ are combined as a diquark, denoted by $q_{[2,3]}$.
Explicitly,
the baryon bound state can be written as~\cite{Dosch:1988hu}
\begin{eqnarray}\label{wf1}
&&
|{\bf B}(P,S,S_z)\rangle=\int\{d^{3}p_{1}\}\{d^{3}p_{2}\}\nonumber\\
&&\times 2(2\pi)^{3}\delta^{3}(\tilde{P}-\tilde{p}_{1}-\tilde{p}_{2})
\sum_{\lambda_{1},\lambda_{2}}
\Psi^{SS_{z}}(\tilde{p}_{1},\tilde{p}_{2},\lambda_{1},\lambda_{2})
|q_1(p_1,\lambda_{1})q_{[2,3]}(p_{2},\lambda_{2})\rangle\,,
\end{eqnarray}
where $p_i$ and $\lambda_i$ stand for the momentum and helicity state, respectively,
and $\Psi^{SS_{z}}(\tilde{p}_{1},\tilde{p}_{2},\lambda_{1},\lambda_{2})$
is the momentum-space wave function. In the light-front frame, one defines 
$P=(P^-,P^+,P_\bot)$ with $P^\pm=P^0\pm P^3$ and $P_\bot=(P^1,P^2)$, and
$p_i=(p_i^-,p_i^+,p_{i\bot})$ with $p_i^\pm=p_i^0\pm p_i^3$ and $p_{i\bot}= (p_i^1, p_i^2)$,
together with $\tilde P=(P^+,P_\bot)$ and $\tilde p_i=(p_i^+, p_{i\bot})$,
which result in
$P^+ P^-=M^2+P_{\bot}^2$ and $p_i^+ p_i^- = {m_i^2+p_{i\bot}^2}$
with $(m_1,m_2)=(m_{q_1},m_{q_2}+m_{q_3})$.
Moreover, $P$ and $p_i$ are related as
$P^{+}=p^+_1+p^+_2$ and $P_{\bot}=p_{1\bot}+p_{2\bot}$,
where
\begin{eqnarray}\label{para1}
&&
p^+_1=(1-x) P^+\,,\; 
p^+_2=x P^+\,,\;\nonumber\\
&&
p_{1\bot}=(1-x) P_\bot-k_\bot\,,\;
p_{2\bot}=xP_\bot+k_\bot\,,
\end{eqnarray}
with $k_\perp$ from $\vec{k}=(k_\perp,k_z)$ the relative momentum.
By means of $e_i\equiv\sqrt{m^2_{i}+\vec{k}^2}$ the energy of the (di)quark 
and $M_0\equiv e_1+e_2$, the above parameters can be rewritten as
\begin{eqnarray}
&&
(x,1-x)=(e_2-k_z,e_1+k_z)/(e_1+e_2)\,,\; 
k_z=\frac{xM_0}{2}-\frac{m^2_{2}+k^2_{\perp}}{2xM_0}\,.
\end{eqnarray}
In addition, we obtain
$M_0^2=(m_{1}^2+k_\bot^2)/(1-x)+(m_{2}^2+k_\bot^2)/x$.
We also get $(\bar P_\mu \gamma^\mu-M_0)u(\bar{P},S_{z})=0$ with $\bar P\equiv p_1+p_2$, 
where $p_{1,2}$ describe the internal motions of the internal quarks. 
Under the Melosh transformation~\cite{Melosh:1974cu}, 
we derive $\Psi^{SS_{z}}$ as~\cite{Ke:2012wa,Ke:2017eqo,Zhao:2018mrg,Hu:2020mxk}
\begin{eqnarray}\label{wf2}
\Psi^{SS_{z}}(\tilde{p}_{1},\tilde{p}_{2},\lambda_{1},\lambda_{2})&=&
\sqrt{\frac{C}{2(p_{1}\cdot\bar{P}+m_{1}M_{0})}}\;\bar{u}(p_{1},\lambda_{1})
\Gamma u(\bar{P},S_{z})\phi(x,k_{\perp})\,,
\end{eqnarray}
where $\Gamma=\Gamma_S(\Gamma_{A}^{(\alpha)})$ 
represents the vertex function 
for the scalar (axial-vector) quantity of the diquark,
given by~\cite{Ke:2012wa,Ke:2017eqo,Zhao:2018mrg,Hu:2020mxk}
\begin{eqnarray}
&&
\Gamma_S=1\,,\;\nonumber\\
&&
\Gamma_{A}=-\frac{1}{\sqrt{3}}\gamma_{5} \strich\epsilon^{*}(p_{2},\lambda_{2})\,,\;
\Gamma_{A}^{\alpha}=\epsilon^{*\alpha}(p_{2},\lambda_{2})\,.
\end{eqnarray}
Moreover, the parameter $C$ for 
$(\Gamma_{S(A)},\Gamma_A^\alpha)$ is given by
\begin{eqnarray}
C=\bigg(
\frac{3(m_{1}M_{0}+p_{1}\cdot\bar{P})}{3m_{1}M_{0}+p_{1}\cdot\bar{P}+
2(p_{1}\cdot p_{2})(p_{2}\cdot\bar{P})/m_{2}^{2}}\,,\;
\frac{3m_{2}^{2}M_{0}^{2}}{2m_{2}^{2}M_{0}^{2}+(p_{2}\cdot\bar{P})^{2}}
\bigg)\,.
\end{eqnarray}
In Eq.~(\ref{wf2}), $\phi(x,k_{\perp})$ is the wave function
that illustrates the momentum distribution of the constituent quark-diquark states.
Here, we present $\phi(x,k_{\perp})$ in the Gaussian form~\cite{Hsiao:2020gtc,Zhao:2018zcb,Ke:2012wa,
Ke:2017eqo,Zhao:2018mrg,Hu:2020mxk,Ke:2019smy}: 
\begin{eqnarray}\label{wf3}
\phi(x,k_{\perp})&=&4\left(\frac{\pi}{\beta^{2}}\right)^{3/4}\sqrt{\frac{e_{1}e_{2}}{x(1-x)M_{0}}}\exp
\left(\frac{-\vec{k}^{2}}{2\beta^{2}}\right)\,,
\end{eqnarray}
with $\beta\equiv\beta_{b[ss]}(\beta_{s[ss]})$ to shape the momentum distribution
of the $b$-$[ss]$ ($s$-$[ss]$) system in the $\Omega_b$ ($\Omega$) bound state.

Using the bound states of
$|\Omega_b(P,S,S_z)\rangle$ and $|\Omega(P,'S',S'_z)\rangle$ 
in Eq.~(\ref{wf1}) and the above identities,
we derive the matrix elements of the $\Omega_b\to \Omega$ transition 
in the light-front frame, 
 given by~\cite{Zhao:2018mrg}
\begin{eqnarray}\label{transitionVA2} 
\langle \bar T^\mu\rangle
&\equiv& 
\langle \Omega(P^{\,\prime},S'=3/2,S_z^\prime)|\bar s\gamma^\mu
(1-\gamma_5)b|\Omega_b(P,S=1/2,S_z)\rangle\nonumber \\
&=&
\int\{d^{3}p_{2}\}\hat C^{-1/2} \phi^{\prime}(x^{\prime},k_{\perp}^{\prime})\phi(x,k_{\perp})
\nonumber \\
&& 
\times\sum_{\lambda_{2}}\bar{u}_{\alpha}(\bar{P}^{\,\prime},S_{z}^{\,\prime})
\left[\bar{\Gamma}^{\,\prime\alpha}_{A}(\strich p_{1}^{\prime}+m_{1}^{\prime})
\gamma^{\mu}(1-\gamma_{5})(\strich p_{1}+m_{1})\Gamma_{A}\right]u(\bar{P},S_{z})\,,
\end{eqnarray}
where $m_1^{(\prime)}=m_{b(s)}$, 
$\bar \Gamma=\gamma^0 \Gamma^\dagger\gamma^0$ and
$\hat C=4p_{1}^{+}p_{1}^{\prime+}(p_{1}
\cdot\bar{P}+m_{1}M_{0})(p_{1}^{\prime}\cdot\bar{P}^{\,\prime}
+m_{1}^{\prime}M_{0}^{\prime})$. 

To determine $F_i^{V,A}$, the identities
$J_{(5)}^{\mu}\equiv\bar{u}\Gamma^{\mu\beta}(\gamma_5) u_{\beta}$ and
$\bar J_{(5)}^{\mu}\equiv\bar{u}\bar \Gamma^{\mu\beta}(\gamma_5) u_{\beta}$
can be useful, where 
$\Gamma^{\mu\beta}=
(\gamma^{\mu}P^{\beta},P^{\,\prime\mu}P^{\beta},P^{\mu}P^{\beta},g^{\mu\beta})$ and
$\bar \Gamma^{\mu\beta}=
(\gamma^{\mu}\bar P^{\beta},\bar P^{\,\prime\mu}\bar P^{\beta},\bar P^{\mu}\bar P^{\beta},g^{\mu\beta})$.
We can hence perform the following calculations~\cite{Zhao:2018mrg,Hsiao:2020gtc},
\begin{eqnarray}\label{F5j}
J_5\cdot \langle T\rangle
&=&Tr\bigg\{
u_{\beta}\bar{u}_{\alpha}
\left[\frac{P^{\alpha}}{M}\left(\gamma^{\mu}F^V_{1}
+\frac{P^{\mu}}{M} F^V_{2}
+\frac{P^{\,\prime\mu}}{M^{\prime}}F^V_{3}\right)+g^{\alpha\mu}F^V_{4}\right]
\gamma_{5}\bar u{\Gamma}_{\mu}^{\beta}\gamma_5\bigg\}\,,
\nonumber\\
\bar J_{5}\cdot \langle \bar T\rangle&=&
\int\{d^{3}p_{2}\}
\hat C^{-1/2}\phi^{\prime}(x^{\prime},k_{\perp}^{\prime})\phi(x,k_{\perp})
\nonumber\\
&\times&
\sum_{\lambda_{2}}
Tr\bigg\{u_{\beta}\bar{u}_{\alpha}
\left[\bar{\Gamma}^{\,\prime\alpha}_{A}(\strich p_{1}^{\prime}+m_{1}^{\prime})
\gamma^{\mu}(\strich p_{1}+m_{1})\Gamma_{A}\right]u\bar{\Gamma}_{\mu}^{\beta}\gamma_5\bigg\}\,.
\end{eqnarray}
By connecting $J_5\cdot \langle T\rangle$ to $\bar J_5\cdot \langle \bar T\rangle$,
that is, $J_5\cdot \langle T\rangle=\bar J_5\cdot \langle \bar T\rangle$, 
$F_i^{V}$ in $J_5\cdot \langle T\rangle$ can be extracted with
$\bar J_5\cdot \langle \bar T\rangle$ in the light-front quark model,
as the other extractions of 
the ${\bf B}_{b(c)}\to {\bf B}^{(*)}$ transition form factors~\cite{Hsiao:2020gtc,
Zhao:2018zcb,Ke:2012wa,Ke:2017eqo,Zhao:2018mrg,Hu:2020mxk,Ke:2019smy}.
Similarly, $J\cdot \langle T\rangle=\bar J\cdot \langle \bar T\rangle$ 
enables us to get $F_i^A$. We will present our results in the next section.

\section{Numerical analysis}
For the numerical analysis, the CKM matrix elements and
the mass (decay constant) of the $J/\Psi$ meson state
are given by~\cite{pdg}
\begin{eqnarray}
(V_{cb},V_{cs})&=&(A\lambda^2,1-\lambda^2/2)\,,\nonumber\\
(m_{J/\Psi},f_{J/\Psi})&=&(3.097,0.418)~\text{GeV}\,,
\end{eqnarray}
with $\lambda=0.2265$ and $A=0.790$ in the Wolfenstein parameterization.
The effective Wilson coefficients $(c^{eff}_1,\,c^{eff}_2)=(1.168,-0.365)$
come from Refs.~\cite{ali,Hsiao:2014mua}. 
In the generalized version of the factorization approach,
$N_c$ is taken as a floating number, in order that 
the non-factorizable effects from QCD corrections can be estimated.
By adopting $N_c=2.15\pm 0.17$ in~\cite{Hsiao:2015cda,Hsiao:2015txa}, 
we obtain $a_2=0.18^{+0.05}_{-0.04}$,
which has been used to interpret ${\cal B}(\Lambda_b\to \Lambda J/\psi)$ 
and ${\cal B}(\Xi_b^-\to \Xi^- J/\psi)$. 

\begin{figure}[t!]
\centering
\includegraphics[width=2.6 in]{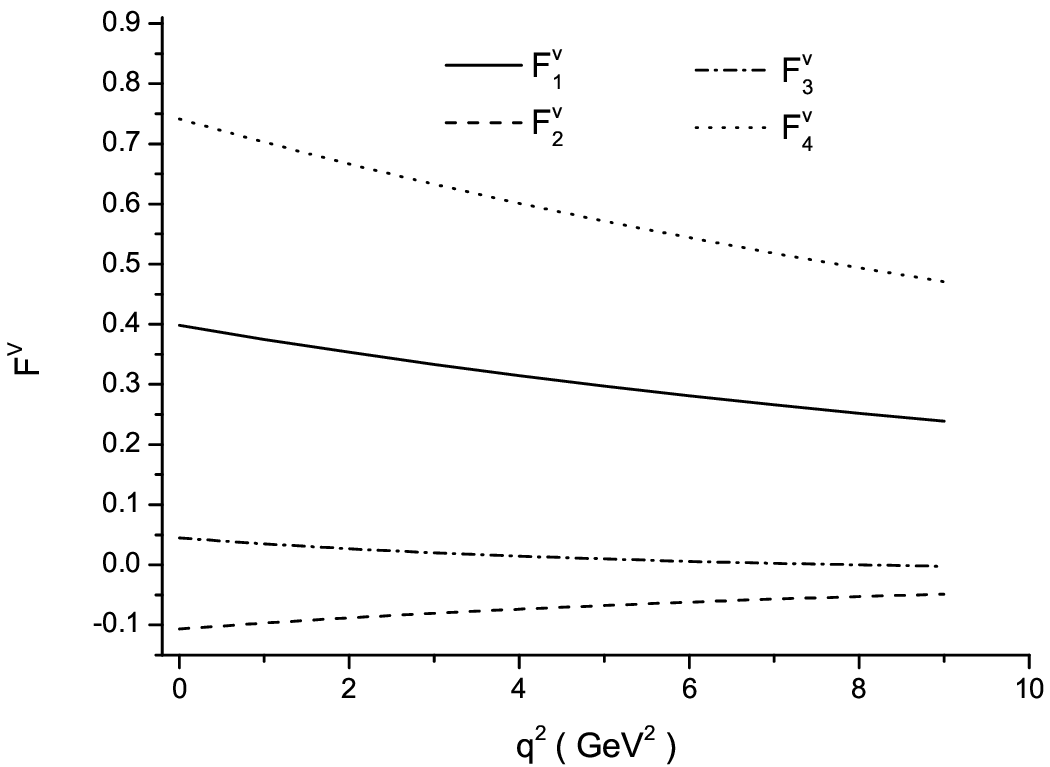}
\includegraphics[width=2.6 in]{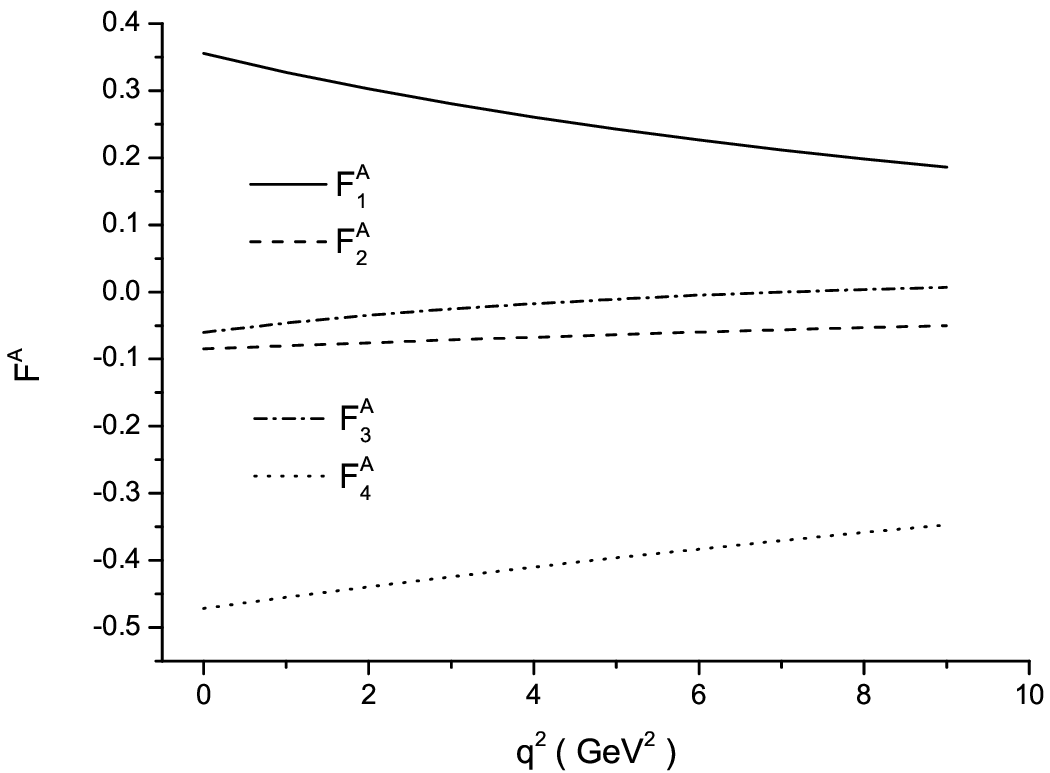}
\caption{$F^{V(A)}_i$ versus $q^2$ ($i=1,2,3,4$).}\label{fig2}
\end{figure}

In terms of $J_5\cdot \langle T\rangle=\bar J_5\cdot \langle \bar T\rangle$
and $J\cdot \langle T\rangle=\bar J\cdot \langle \bar T\rangle$
and the theoretical inputs in Eqs.~(\ref{wf3}, \ref{transitionVA2}, \ref{F5j}),
given by~\cite{Ke:2019smy}
\begin{eqnarray}\label{mqbetaq}
(m_b,\beta_{b[ss]})&=&(5.00\pm 0.20,0.78\pm 0.04)~\mathrm{GeV}\,,\nonumber\\ 
(m_s,\beta_{s[ss]})&=&(0.38,0.48)~\mathrm{GeV}\,,
\end{eqnarray}
we derive $F^{V(A)}_i$ as the functions of $q^2$, depicted in Fig.~\ref{fig2}.
It is common that one parameterizes the form factors 
in the dipole expressions~\cite{Cheng:2003sm,Choi:2013ira,Hsiao:2019wyd},
which reproduce the momentum dependences derived in the quark model.
Subsequently, the form factors can have simple forms to be used in the weak decays.
In our case, we present~\cite{Zhao:2018zcb,Hsiao:2020gtc}
\begin{eqnarray}\label{eq:LFparameters}
F(q^2)=\frac{F(0)}{1-a\left(q^2/m_F^2\right)+b\left(q^4/m_F^4\right)}\,,
\end{eqnarray}
with $m_F$, $a$, $b$ and $F(0)$ at $q^2=0$ given in Table~\ref{tab1},
in order to describe the momentum behaviors of $F_{i}^{V,A}$ in Fig.~\ref{fig2}.

%
%
%
\begin{table}[b!]
\caption{The $\Omega_b\to\Omega$ transition form factors
with $(F(0),a,b)$ in Eq.~(\ref{eq:LFparameters}), where $m_F=6.05$~GeV
is from $m_{\Omega_b}$. 
The uncertainties come from $m_b$ and $\beta_{b[ss]}$ in Eq.~(\ref{mqbetaq}).}\label{tab1}
{
\begin{tabular}{|c|r|r|r|} \hline
    & $F(0)\;\;\;\;\;\;$ & $a\;\;\;$ & $b\;\;\;$
\\ \hline \hline
$F^V_{1}$ & $0.371^{+0.045}_{-0.042}$ & $-2.22$ & $ 2.37$
\\ 
$F^V_{2}$ & $-0.104^{+0.022}_{-0.025}$ & $-3.19$ & $4.69$
\\ 
$F^V_{3}$ & $0.040^{+0.042}_{-0.035}$ & $4.11$ & $11.38$
\\ 
$F^V_{4}$ & $0.692^{+0.054}_{-0.051}$ & $-2.05$ & $1.91$
\\ \hline
\end{tabular}
\begin{tabular}{|c|r|r|r|} \hline
& $F(0)\;\;\;\;\;\;$ & $a\;\;\;$ & $b\;\;\;$
\\ \hline \hline
$F^A_{1}$ & $0.329^{+0.121}_{-0.110}$ & $-1.93$ & $2.73$
\\ 
$F^A_{2}$ & $-0.081^{+0.022}_{-0.020}$ & $-3.31$ & $4.36$
\\ 
$F^A_{3}$ & $-0.064^{+0.130}_{-0.140}$ & $-3.16$ & $0.77$
\\ 
$F^A_{4}$ & $-0.416^{+0.092}_{-0.082}$ & $-1.89$ & $0.99$
\\ \hline
\end{tabular}}
\end{table}
%

Thus, we calculate the branching fraction and fragmentation fraction as
\begin{eqnarray}\label{result1}
&&
{\cal B}(\Omega_b^-\to\Omega^- J/\Psi)=(5.3^{+3.3+3.8}_{-2.1-2.7})\times 10^{-4}\,,\nonumber\\
&&
f_{\Omega_b}=(0.54^{+0.34+0.39+0.21}_{-0.22-0.28-0.15})\times 10^{-2}\,,
\end{eqnarray}
where $f_{\Omega_b}$ is extracted with ${\cal B}(\Omega_b^-\to\Omega^- J/\Psi)$
and the data in Eq.~(\ref{data1}). Moreover,
the first and second uncertainties come from $a_2$ and $F_i^{V,A}$, respectively,
and the third one for $f_{\Omega_b}$ is from the measurement.

\section{Discussions and Conclusions}
Because of the insufficient information on the $\Omega_b\to {\bf B}^*$ transition,
the $\Omega_b$ decays have not been richly explored.
In the light-front quark model, we calculate 
the $\Omega_b\to \Omega$ transition form factors. We can hence predict 
${\cal B}(\Omega_b\to\Omega J/\Psi)=(5.3^{+3.3+3.8}_{-2.1-2.7})\times 10^{-4}$,
which is compatible with those of the anti-triplet $b$-baryon decays
${\cal B}(\Lambda_b\to\Lambda J/\Psi)=(3.3\pm 2.0)\times 10^{-4}$ and 
${\cal B}(\Xi_b^-\to\Xi^- J/\Psi)=(5.1\pm 3.2)\times 10^{-4}$~\cite{Hsiao:2015cda,Hsiao:2015txa}.
On the other hand,
${\cal B}(\Omega_b\to\Omega J/\Psi)=8.1\times 10^{-4}$ 
is given by the authors of Ref.~\cite{Gutsche:2018utw}.
In addition, the total decay width $\Gamma(\Omega_b\to \Omega J/\Psi)
=3.15 a_2^2\times 10^{10}$~s$^{-1}$~\cite{Cheng:1996cs}
leads to ${\cal B}(\Omega_b\to\Omega J/\Psi)=16.7\times 10^{-4}$, 
where we have used $a_2=0.18$ for the demonstration.

In the helicity basis, the branching fraction is given by
\begin{eqnarray}
{\cal B}\propto (|{\cal H}_V|^2+|{\cal H}_A|^2)\,,
\end{eqnarray}
where $|{\cal H}_{V(A)}|^2\equiv
|H_{\frac321}^{V(A)}|^2+|H_{\frac121}^{V(A)}|^2+|H_{\frac120}^{V(A)}|^2$.
It is found that $(|{\cal H}_{V}|^2,|{\cal H}_{A}|^2)$ give (19,81)\% of ${\cal B}$;
besides, 
$(|H_{\frac321}^{A}|^2,|H_{\frac121}^{A}|^2,|H_{\frac120}^{A}|^2)/|{\cal H}_{A}|^2
=(54.0,22.4,23.6)\%$, such that $F_4^A$ gives the main contribution
to ${\cal B}(\Omega_b\to\Omega J/\Psi)$.

In Eq.~(\ref{result1}), $f_{\Omega_b}=0.54\times 10^{-2}$
agrees with the previous upper limit of 0.108~\cite{Hsiao:2015txa}.
By comparing our extraction to $f_{\Lambda_b}= 0.175\pm 0.106$ 
and $f_{\Xi_b}= 0.019\pm 0.013$~\cite{Hsiao:2015txa},
it demonstrates that the $b$ to $\Omega_b$ productions
are much more difficult than the $b$ to ${\bf B}_b$ ones.
Since the fragmentation fraction has not been determined experimentally,
the branching fractions of the $\Omega_b$ decays should be partially measured 
with the factor $f_{\Omega_b}$. Therefore,
our extraction for $f_{\Omega_b}$ can be useful.
With $f_{\Omega_b}$ of Eq.~(\ref{result1}) added to the branching fractions, one
can compare his theoretical results to future measurements of the $\Omega_b$ decays.

In summary, we have investigated the charmful $\Omega_b$ decay channel
$\Omega_b^-\to \Omega^- J/\Psi$.
In the light-front quark model, we have studied
the $\Omega_b\to\Omega$ transition form factors $(F_i^V,F_i^A)$ ($i=1,2, ...,4$).
We have hence predicted
${\cal B}(\Omega_b^-\to\Omega^- J/\Psi)=(5.3^{+3.3+3.8}_{-2.1-2.7})\times 10^{-4}$,
which is compatible with those of 
the $\Lambda_b\to\Lambda J/\Psi$ and $\Xi_b^-\to\Xi^- J/\Psi$ decays.
In addition, $F_4^A$ has been found to give the main contribution. Particularly, 
we have extracted $f_{\Omega_b}=(0.54^{+0.34+0.39+0.21}_{-0.22-0.28-0.15})\times 10^{-2}$
from the partial observation 
$f_{\Omega_b}{\cal B}(\Omega_b^-\to \Omega^- J/\Psi)=(2.9^{+1.1}_{-0.8})\times 10^{-6}$.
Since $f_{\Omega_b}$ has not been determined experimentally,
by adding $f_{\Omega_b}$ to the branching fractions, 
one is allowed to compare his calculations to future observations
of the $\Omega_b$ decays.

\newpage
\section*{ACKNOWLEDGMENTS}
YKH was supported in part 
by National Science Foundation of China (Grants No.~11675030 and No.~12175128).
CCL was supported in part by CTUST (Grant No. CTU109-P-108).

\end{document}